\documentclass{elsarticle}
\usepackage{lipsum}
\makeatletter
\def\ps@pprintTitle{%
 \let\@oddhead\@empty
 \let\@evenhead\@empty
 \def\@oddfoot{}%
 \let\@evenfoot\@oddfoot}

\usepackage{natbib,amssymb,amsbsy,fleqn}

\usepackage{chemformula}
\usepackage{tikz}
\usepackage{float}
\usepackage{xcolor,colortbl}
\usepackage{braket}
\usepackage{amsfonts}
\usepackage{graphicx}
\usepackage{amsthm}
\usepackage{amssymb}
\usepackage{amsmath}
\usepackage{mathrsfs}
\usepackage{fancybox}
\usepackage{fancyhdr}
\usepackage{subfigure}

\usepackage{color, colortbl}
\definecolor{Cyan}{rgb}{0.88,1,1}
\definecolor{Orchid}{rgb}{0.6, 0.2, 0.8}
\definecolor{darkpink}{rgb}{0.91, 0.33, 0.5}
\definecolor{Gray}{gray}{0.8}

\begin{document}

\title{ Quantum theory of damping: A new approach to temporal evolution of density operator without the rotating wave approximation}

\author{A. Marashi Nasab, E. Ghasemian and M. K. Tavassoly$^*$\\
 \footnotesize{ Atomic and Molecular Group, Faculty of Physics, Yazd University, Yazd, Iran} \\
  \footnotesize{$^*$  E-mail: mktavassoly@yazd.ac.ir}}

\date{\today}

\begin{abstract}
	The Lindblad master equation is the standard framework for describing dissipative dynamics of the open quantum system context. Its conventional derivation, however, relies on the rotating-wave approximation (RWA), which neglects the counter rotating terms in the system-reservoir interaction. Here, we derive a generalized Lindblad master equation beyond the RWA within the density operator formalism, retaining the full system-reservoir coupling. We consider two prototypical open quantum systems, a two-level atom and a single-mode quantized field, each interacting with thermal or squeezed reservoirs described by the Wigner-Weisskopf theory. The resulting master equations are applied to the evolution of representative physical observables and compared with the conventional RWA based formulation. We show that counter-rotating terms produce significant modifications to the dissipative dynamics and therefore cannot, in general, be neglected. For a thermal reservoir, increasing the mean thermal photon number, $n_{\mathrm{th}}$, enhances the decay of $\langle\hat{\sigma}{_y}\rangle$ beyond the RWA but suppresses it within the RWA. Likewise, the evolution of $\langle\hat{\sigma}{_z}\rangle$ exhibits opposite damping behavior with increasing $\Gamma t$ and squeezing parameter $r$. These findings demonstrate the importance of including counter rotating terms for an accurate description of open quantum systems.
\end{abstract}
\maketitle

\section{Introduction}
Quantum systems are generally classified as either closed or open, depending on whether they interact with their environment. A closed quantum system evolves unitarily according to the time-dependent Schrödinger equation, whereas an open quantum system undergoes decoherence and dissipation due to its coupling with the surroundings. Since complete isolation is rarely achievable, the theory of open quantum systems has become indispensable for describing realistic quantum phenomena and forms the theoretical basis of quantum optics, quantum information processing, quantum metrology, and quantum technologies.
The dynamics of Markovian open quantum systems are commonly described by the Lindblad master equation \cite{Scully,Knight,Agarwal,Albert,Kosov}.
A key assumption underlying the conventional derivation of the Lindblad equation is the rotating wave approximation (RWA), through which rapidly oscillating counter rotating terms (CRTs) in the system-reservoir interaction are discarded. Although this approximation greatly simplifies the derivation, its validity is restricted to weak coupling and well-separated time scales. Recent advances in cavity QED, superconducting circuits, semiconductor microcavities, and other light-matter platforms have enabled strong and ultrastrong coupling regimes, where the neglected CRTs give rise to measurable phenomena such as virtual-photon processes, vacuum fluctuations, and energy-level shifts. In this line, applying RWA in quantum systems causes significant effects in the transient dynamics \cite{Balzer,Thingna,Damanet} and  transport properties \cite{Wichterich}.
In a series of papers Seke  paid his attention to the influence of CRTs on the (multi)atom-field interaction with different initial atom and field conditions \cite{Seke1,Seke2,Seke3}. 
Consequently, the validity of the standard Lindblad equation in these regimes has become an important theoretical question. Becker \textit{et al.} \cite{Becker}, for example, proposed an alternative Lindbladian approximation that remains accurate over a broader parameter range, highlighting the need for improved microscopic descriptions of dissipation beyond the conventional RWA.
Besides the Lindblad formalism, dissipative quantum dynamics have also been investigated using quantum jump methods \cite{Dehmelt,Knight1,Benatti} and phenomenological non-Hermitian Hamiltonians \cite{Vogel,Nori,Dong,Eftekhari}. Although these approaches successfully describe many cavity QED and optomechanical phenomena, they generally rely on phenomenological assumptions or still employ the RWA. Likewise, extensive studies have examined atom-field interactions beyond the RWA by explicitly retaining the CRTs in generalized Jaynes-Cummings-type models \cite{Peng1,Peng2,Asili,Rastegarzadeh,Zhang}. These works have demonstrated the importance of CRTs in the dynamics of the considered systems via various methods.\\
Despite the central role of the Lindblad master equation in open quantum theory, a microscopic derivation that explicitly retain the CRTs remains largely unexplored. This omission is particularly significant because the interaction Hamiltonian constitutes the starting point of every microscopic master equation derivation, and neglecting the CRTs may alter both the structure of the dissipator and the resulting system dynamics. Establishing a Lindblad description beyond the RWA is therefore of fundamental interest for accurately modelling dissipative quantum systems operating outside the weak-coupling regime. To the best of our knowledge, such a systematic derivation has not yet been reported.
In this work, we derive a generalized Lindblad master equation without invoking the rotating wave approximation. Starting from the complete atom-reservoir and field-reservoir interaction Hamiltonians, including the counter rotating terms, we obtain the reduced dynamics by tracing over the reservoir degrees of freedom. The formalism is developed for both thermal and squeezed reservoirs \cite{Moya,SHAN} and is applied to dissipative two-level atoms and single-mode cavity fields. By comparing the resulting dynamics with those predicted by the conventional RWA-based master equation, we identify regimes in which the CRTs produce substantial modifications to physically observable quantities. Our results provide a more complete microscopic description of open quantum systems and are particularly relevant to strong and ultra strong light-matter interaction regimes, such as superconducting circuit QED and related quantum optical platforms \cite{Salimian}.\\
The paper is organized as follows. Section 2 presents the derivation of generalized master equation. Sections 3 and 4 discuss atomic dissipation in thermal and squeezed reservoirs, respectively, while sections 5 and 6 consider field dissipation in the corresponding environments. A few applications of the formalism are presented in section 7, and conclusions are given in section 8.
\section{Time evolution of dissipative density operator: General formalism}

The total density operator associated with the system-reservoir is denoted by {$\rho_{_{SR}}$} and
the system-reservoir interaction in the interaction picture is given by {$\hat{V}(t)$}. The equation of motion for {$\hat{\rho}_{_{SR}}$} is therefore given by {$i{\hbar}{\dot{\hat{\rho}}_{_{SR}}}=[\hat{V}(t),{\hat{\rho}_{_{SR}}}(t)].$} 
It is well-known that the time evolution of the reduced density operator of the system, {$\hat\rho_{_{S}}$} can be obtained by tracing over the degrees of freedom of the reservoir. Accordingly, we may have the following relations \cite{Molmer},
\begin{eqnarray}\label{eq:1}
{\dot{\hat{\rho}}_{_{S}}}{(t)}&=&{Tr_{_{R}}[{\dot{\hat{\rho}}_{_{SR}}}(t)]}\nonumber\\
&=&-{(i/{\hbar})}{Tr_{_{R}}[\hat{V}(t),{\hat{\rho}_{_{SR}}}(t')]}-{(1/{\hbar^{2}})}{Tr_{_{R}}}{\int_{t_i}^{t}}{[\hat{V}(t),[\hat{V}(t'),{\hat{\rho}_{_{SR}}}(t')]]}{dt'}.
\end{eqnarray}
In the continuation of this section by considering an atomic and a field system, separately, their interactions with two thermal and squeezed reservoirs as two dissipation sources are evaluated distinctly, by which we obtain the time evolution equations of density operator of the atom and field in each case separately, while the RWA puts aside and CRTs in the encountered Hamiltonians (Figure 1). The atom is a two-level system  and the field is a single-mode field. 
\newpage 
\begin{figure}[!h]
	\centering
	\includegraphics[width=9cm]{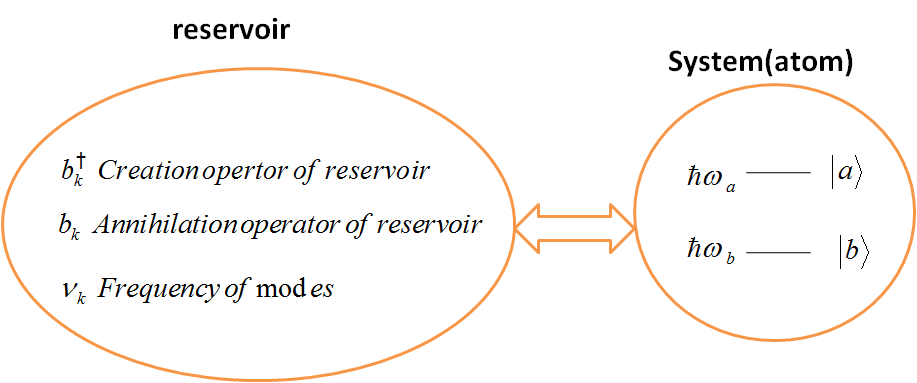}
	\caption{Interaction between a two-level atom and a reservoir.}
\end{figure}
\section{Two-level atomic dissipation in contact with thermal reservoir}
In this section we aim to consider a two-level atomic system interacting with a thermal reservoir.
This model has been frequently introduced  in the literature \cite{Scully}, while in the interaction Hamiltonian the RWA has been applied. Now, as the main purpose of the  paper, we want to consider the same model, however in the presence of CRTs, i.e. without RWA. Accordingly, the atom-reservoir interaction picture Hamiltonian can be written as follows \cite{Zhou,Gonouiezadeh},

\begin{eqnarray}\label{eq:2}
{\hat{V}(t)}&=&{\hbar}{\sum\limits_{k}}{g_{k}}[{\hat{b}^\dagger_{k}{\hat{\sigma}_{-}}}{e^{-i(\omega-\nu_{k})t}}+{\hat{b}_{k}{\hat{\sigma}_{+}}}{e^{i(\omega-\nu_{k})t}}
+{\hat{b}_{k}{\hat{\sigma}_{-}}}{e^{-i(\omega+\nu_{k})t}}\nonumber\\
&+&{\hat{b}^\dagger_{k}{\hat{\sigma}_{+}}}{e^{i(\omega+\nu_{k})t}}],
\end{eqnarray}
where ${g_{k}}$ is the coupling constant of field mode ${k}$ with atom, $\hat{b}_{k}$ and $\hat{b}^\dagger_{k}$ respectively are annihilation and creation reservoir operators of mode $k$, {$\left |a \right \rangle$ } is the exited state, {$\left |b \right \rangle$ } is the ground state, {$\hat{\sigma}_{-}={\left |b \right \rangle }{\left \langle a \right |}$} and {$\hat{\sigma}_{+}={\left |a \right \rangle }{\left \langle b \right |}$} respectively are atomic lowering and raising  operators and {$\omega (\nu_{k})$} is the transition frequency of the atom (frequency of the field mode ${k}$). Recall that usually by considering RWA the last two terms in (\ref{eq:2}) containing the exponential terms of the form $\omega + \nu_k$ have been ignored.
On choosing {$\hat{\rho}_{s}$}={$\hat{\rho}_{at}$} and inserting Eq. (\ref{eq:2}) into (\ref{eq:1}), tracing over the field modes and after a lengthy but straightforward  calculations, the equation governing the time evolution of the atomic density operator has been obtained as
\begin{eqnarray}\label{eq:3}
{{\dot{\hat{\rho}}_{at}}{(t)}}&=&-{i\sum\limits_{k}}{g_{k}}{\langle{\hat{b}_{k}}\rangle}
{[\hat{\sigma}_{+}},{\hat{\rho}_{at}{(t)}]}{e^{i(\omega-\nu_{k})t}}-{i\sum\limits_{k}}{g_{k}}{\langle{\hat{b}_{k}}\rangle}{[\hat{\sigma}_{-}},{\hat{\rho}_{at}{(t)}]}{e^{-i(\omega+\nu_{k})t}}\nonumber\\
&-&{\int\limits_{t_{i}}^{t}}{dt'}{\sum_{k}}{\sum_{k'}}{g_{k}}{g_{k'}}{\langle{\hat{b}_{k}}{\hat{b}_{k'}^{\dag}}\rangle}[{\hat{\sigma}_{+}},{\hat{\sigma}_{-}}{\hat{\rho}_{at}{(t')}}]{e^{i(\omega-\nu_{k})t-i(\omega-\nu_{k'})t'}}\nonumber\\
&+&{\int\limits_{t_{i}}^{t}}{dt'}{\sum_{k}}{\sum_{k'}}{g_{k}}{g_{k'}}{\langle{\hat{b}_{k}}{\hat{b}_{k'}^{\dag}}\rangle}[{\hat{\sigma}_{+}},{\hat{\rho}_{at}{(t')}}{\hat{\sigma}_{-}}]{e^{i(\omega-\nu_{k})t-i(\omega-\nu_{k'})t'}}\nonumber
\end{eqnarray}
\begin{eqnarray}\label{Hintint}
&-&{\int\limits_{t_{i}}^{t}}{dt'}{\sum_{k}}{\sum_{k'}}{g_{k}}{g_{k'}}{\langle{\hat{b}_{k}}{\hat{b}_{k'}}\rangle}[{\hat{\sigma}_{+}},{\hat{\sigma}_{+}}{\hat{\rho}_{at}{(t')}}]{e^{i(\omega-\nu_{k})t+i(\omega-\nu_{k'})t'}}\nonumber\\
&+&{\int\limits_{t_{i}}^{t}}{dt'}{\sum_{k}}{\sum_{k'}}{g_{k}}{g_{k'}}{\langle{\hat{b}_{k}}{\hat{b}_{k'}}\rangle}[{\hat{\sigma}_{+}},{\hat{\rho}_{at}{(t')}}{\hat{\sigma}_{+}}]{e^{i(\omega-\nu_{k})t+i(\omega-\nu_{k'})t'}}\nonumber\\
&-&{\int\limits_{t_{i}}^{t}}{dt'}{\sum_{k}}{\sum_{k'}}{g_{k}}{g_{k'}}{\langle{\hat{b}_{k}}{\hat{b}_{k'}^{\dag}}\rangle}[{\hat{\sigma}_{+}},{\hat{\sigma}_{+}}{\hat{\rho}_{at}{(t')}}]{e^{i(\omega-\nu_{k})t+i(\omega+\nu_{k'})t'}}\nonumber\\
&+&{\int\limits_{t_{i}}^{t}}{dt'}{\sum_{k}}{\sum_{k'}}{g_{k}}{g_{k'}}{\langle{\hat{b}_{k}}{\hat{b}_{k'}^{\dag}}\rangle}[{\hat{\sigma}_{+}},{\hat{\rho}_{at}{(t')}}{\hat{\sigma}_{+}}]{e^{i(\omega-\nu_{k})t+i(\omega+\nu_{k'})t'}}\nonumber\\
&-&{\int\limits_{t_{i}}^{t}}{dt'}{\sum_{k}}{\sum_{k'}}{g_{k}}{g_{k'}}{\langle{\hat{b}_{k}}{\hat{b}_{k'}}\rangle}[{\hat{\sigma}_{+}},{\hat{\sigma}_{-}}{\hat{\rho}_{at}{(t')}}]{e^{i(\omega-\nu_{k})t-i(\omega+\nu_{k'})t'}}\nonumber\\
&+&{\int\limits_{t_{i}}^{t}}{dt'}{\sum_{k}}{\sum_{k'}}{g_{k}}{g_{k'}}{\langle{\hat{b}_{k}}{\hat{b}_{k'}}\rangle}[{\hat{\sigma}_{+}},{\hat{\rho}_{at}{(t')}}{\hat{\sigma}_{-}}]{e^{i(\omega-\nu_{k})t-i(\omega+\nu_{k'})t'}}\nonumber\\
&-&{\int\limits_{t_{i}}^{t}}{dt'}{\sum_{k}}{\sum_{k'}}{g_{k}}{g_{k'}}{\langle{\hat{b}_{k}}{\hat{b}_{k'}^{\dag}}\rangle}[{\hat{\sigma}_{-}},{\hat{\sigma}_{-}}{\hat{\rho}_{at}{(t')}}]{e^{-i(\omega+\nu_{k})t-i(\omega-\nu_{k'})t'}}\nonumber\\
&+&{\int\limits_{t_{i}}^{t}}{dt'}{\sum_{k}}{\sum_{k'}}{g_{k}}{g_{k'}}{\langle{\hat{b}_{k}}{\hat{b}_{k'}^{\dag}}\rangle}[{\hat{\sigma}_{-}},{\hat{\rho}_{at}{(t')}}{\hat{\sigma}_{-}}]{e^{-i(\omega+\nu_{k})t-i(\omega-\nu_{k'})t'}}\nonumber\\
&-&{\int\limits_{t_{i}}^{t}}{dt'}{\sum_{k}}{\sum_{k'}}{g_{k}}{g_{k'}}{\langle{\hat{b}_{k}}{\hat{b}_{k'}}\rangle}[{\hat{\sigma}_{-}},{\hat{\sigma}_{+}}{\hat{\rho}_{at}{(t')}}]{e^{-i(\omega+\nu_{k})t+i(\omega-\nu_{k'})t'}}\nonumber\\
&+&{\int\limits_{t_{i}}^{t}}{dt'}{\sum_{k}}{\sum_{k'}}{g_{k}}{g_{k'}}{\langle{\hat{b}_{k}}{\hat{b}_{k'}}\rangle}[{\hat{\sigma}_{-}},{\hat{\rho}_{at}{(t')}}{\hat{\sigma}_{+}}]{e^{-i(\omega+\nu_{k})t+i(\omega-\nu_{k'})t'}}\nonumber\\
&-&{\int\limits_{t_{i}}^{t}}{dt'}{\sum_{k}}{\sum_{k'}}{g_{k}}{g_{k'}}{\langle{\hat{b}_{k}}{\hat{b}_{k'}^{\dag}}\rangle}[{\hat{\sigma}_{-}},{\hat{\sigma}_{+}}{\hat{\rho}_{at}{(t')}}]{e^{-i(\omega+\nu_{k})t+i(\omega+\nu_{k'})t'}}\nonumber\\
&+&{\int\limits_{t_{i}}^{t}}{dt'}{\sum_{k}}{\sum_{k'}}{g_{k}}{g_{k'}}{\langle{\hat{b}_{k}}{\hat{b}_{k'}^{\dag}}\rangle}[{\hat{\sigma}_{-}},{\hat{\rho}_{at}{(t')}}{\hat{\sigma}_{+}}]{e^{-i(\omega+\nu_{k})t+i(\omega+\nu_{k'})t'}}\nonumber\\
&-&{\int\limits_{t_{i}}^{t}}{dt'}{\sum_{k}}{\sum_{k'}}{g_{k}}{g_{k'}}{\langle{\hat{b}_{k}}{\hat{b}_{k'}}\rangle}[{\hat{\sigma}_{-}},{\hat{\sigma}_{-}}{\hat{\rho}_{at}{(t')}}]{e^{-i(\omega+\nu_{k})t-i(\omega+\nu_{k'})t'}}\nonumber\\
&+&{\int\limits_{t_{i}}^{t}}{dt'}{\sum_{k}}{\sum_{k'}}{g_{k}}{g_{k'}}{\langle{\hat{b}_{k}}{\hat{b}_{k'}}\rangle}[{\hat{\sigma}_{-}},{\hat{\rho}_{at}{(t')}}{\hat{\sigma}_{-}}]{e^{-i(\omega+\nu_{k})t-i(\omega+\nu_{k'})t'}}+H.c.
\end{eqnarray}

In this line, it can be easily shown that for thermal reservoir one can arrive at,
\begin{eqnarray}\label{eq:4}
&&{\langle{\hat{b}_{k}}\rangle}_{th}={\langle{\hat{b}_{k}^{\dag}}\rangle}_{th}=0,
~~~~~~~~~~~~~~~~~~~{\langle{\hat{b}_{k}^{\dag}}{\hat{b}_{k'}}\rangle}_{th}={\bar{n}_{k}\delta_{kk'}},
\nonumber\\
&&{\langle{\hat{b}_{k}}{\hat{b}_{k'}^{\dag}}\rangle}_{th}=({\bar{n}_{k}+1)\delta_{kk'}},
~~~~~~~~~~~~{\langle{\hat{b}_{k}^{\dag}}{\hat{b}_{k'}^{\dag}}\rangle}_{th}={\langle{\hat{b}_{k}}{\hat{b}_{k'}}\rangle}_{th}=0,
\end{eqnarray}
where $\bar{n}_k$ is the mean photon number in the mode $k$ of the thermal reservoir. Via substituting of the expectation values introduced in (\ref{eq:4}) into (\ref{eq:3}), one obtain as, 
\begin{eqnarray}\label{eq:5}
{{\dot{\hat{\rho}}_{at}}{(t)}}&=&-{\int\limits_{t_{i}}^{t}}{dt'}{\sum_{k}}{g_{k}^{2}}{(\bar{n}_{th}+1)}[{\hat{\sigma}_{+}},{\hat{\sigma}_{-}}{\hat{\rho}_{at}{(t')}}]{e^{+i(\omega-\nu_{k})t-i(\omega-\nu_{k})t'}}\nonumber\\
&+&{\int\limits_{t_{i}}^{t}}{dt'}{\sum_{k}}{g_{k}^{2}}{(\bar{n}_{th}+1)}[{\hat{\sigma}_{+}},{\hat{\rho}_{at}{(t')}}{\hat{\sigma}_{-}}]{e^{+i(\omega-\nu_{k})t-i(\omega-\nu_{k})t'}}\nonumber\\
&-&{\int\limits_{t_{i}}^{t}}{dt'}{\sum_{k}}{g_{k}^{2}}{(\bar{n}_{th}+1)}[{\hat{\sigma}_{+}},{\hat{\sigma}_{+}}{\hat{\rho}_{at}{(t')}}]{e^{+i(\omega-\nu_{k})t+i(\omega+\nu_{k})t'}}\nonumber\\
&+&{\int\limits_{t_{i}}^{t}}{dt'}{\sum_{k}}{g_{k}^{2}}{(\bar{n}_{th}+1)}[{\hat{\sigma}_{+}},{\hat{\rho}_{at}{(t')}}{\hat{\sigma}_{+}}]{e^{+i(\omega-\nu_{k})t+i(\omega+\nu_{k})t'}}\nonumber\\
&-&{\int\limits_{t_{i}}^{t}}{dt'}{\sum_{k}}{g_{k}^{2}}{(\bar{n}_{th}+1)}[{\hat{\sigma}_{-}},{\hat{\sigma}_{-}}{\hat{\rho}_{at}{(t')}}]{e^{-i(\omega+\nu_{k})t-i(\omega-\nu_{k})t'}}\nonumber\\
&+&{\int\limits_{t_{i}}^{t}}{dt'}{\sum_{k}}{g_{k}^{2}}{(\bar{n}_{th}+1)}[{\hat{\sigma}_{-}},{\hat{\rho}_{at}{(t')}}{\hat{\sigma}_{-}}]{e^{-i(\omega+\nu_{k})t-i(\omega-\nu_{k})t'}}\nonumber\\
&-&{\int\limits_{t_{i}}^{t}}{dt'}{\sum_{k}}{g_{k}^{2}}{(\bar{n}_{th}+1)}[{\hat{\sigma}_{-}},{\hat{\sigma}_{+}}{\hat{\rho}_{at}{(t')}}]{e^{-i(\omega+\nu_{k})t+i(\omega+\nu_{k})t'}}\nonumber\\
&+&{\int\limits_{t_{i}}^{t}}{dt'}{\sum_{k}}{g_{k}^{2}}{(\bar{n}_{th}+1)}[{\hat{\sigma}_{-}},{\hat{\rho}_{at}{(t')}}{\hat{\sigma}_{+}}]{e^{-i(\omega+\nu_{k})t+i(\omega+\nu_{k})t'}}+H.c.
\end{eqnarray}
To go proceed with the help of the Weisskopf-Wigner approximation, as is well known, the summation over $k$ may be replaced by an integral of the form, i.e., 
{${\sum_{k}}{\rightarrow}\frac{2V}{{(2\pi )}^{3}}{\int\limits_{0}^{2\pi}}{d\phi}{\int\limits_{0}^{\pi}}{d\theta}~{\sin{\theta}}{\int\limits_{0}^{\infty}}{dk}~{k^{2}}$},
from which one can solve the appeared typical integrals in (\ref{eq:5}) through the following procedure, 
\begin{eqnarray}\label{eq:6}
{\int\limits_{t_{i}}^{t}}{dt'}{\sum_{k}}{g_{k}^{2}}{e^{-i(\omega+\nu_{k})t+i(\omega+\nu_{k})t'}}&=&\frac{2V}{{(2\pi )}^{3}}{\int\limits_{t_{i}}^{t}}{dt'}{\int\limits_{0}^{2\pi}}{d\phi}{\int\limits_{0}^{\pi}}{d\theta}~{\sin{\theta}}{\int\limits_{0}^{\infty}}{dk}~{k^{2}}{g_{k}^{2}}\nonumber\\
&\times&{e^{-i(\omega+\nu_{k})t+i(\omega+\nu_{k})t'}}.
\end{eqnarray}
In addition, using the relations,
\begin{eqnarray}\label{eq:7}
{\int\limits_{-\infty}^{\infty}}{d\nu_{k}}{e^{-i(\omega+\nu_{k})(t-t')}}={2\pi}{\delta{(t-t')}},~~~{g_{k}^{2}}=\frac{Q_{ab}^{2}{\nu_{k}}{{\cos^{2}\theta}}}{2{\hbar}{\epsilon_{0}}{V}},
\end{eqnarray}
wherein {$Q_{ab}$} is the electric dipole moment of the atom, {$V$} is the quantization interaction volume and {$\theta$}  is the polar angle between {$Q_{ab}$} and {$z$}-axis and substituting into (\ref{eq:6}) one finally deduces, 
\begin{eqnarray}\label{eq:8}
{\int\limits_{t_{i}}^{t}}{dt'}{\sum_{k}}{g_{k}^{2}}{e^{-i(\omega+\nu_{k})t+i(\omega+\nu_{k})t'}}={\frac{1}{{{4\pi}{\epsilon_{0}}}}{\frac{2{\omega}^{3}{Q_{ab}^{2}}}{3{\hbar}{c^{3}}}}}=\frac{\Gamma}{2}.
\end{eqnarray}
Utilizing the above definitions, parallel to calculating the commutation relations appeared in Eq. (\ref{eq:5}), we finally arrive at, 
\begin{eqnarray}\label{eq:9}
{\dot{\hat{\rho}}_{at}}{(t)}={\dot{\hat{\rho}}^{\tiny{\mbox{RWA}}}_{at}}{(t)}+{\dot{\hat{\rho}}^{\tiny{\mbox{WRWA}}}_{at}}{(t)},
\end{eqnarray}
wherein the first and second terms are respectively associated with the time evolution of the atom density operator with and without {\small{\mbox{RWA}}}, are defined as below,
\begin{eqnarray}\label{eq:10}
{\dot{\hat{\rho}}^{\tiny{\mbox{RWA}}}_{at}}{(t)}&=&{-\bar{n}_{th}}({\Gamma/{2}})({{\hat{\rho}}_{at}}{\hat{{\sigma}_{-}}{\hat{\sigma}_{+}}{-}{\hat{\sigma}_{+}}{\hat{\rho}}_{at}}{\hat{\sigma}_{-}})\nonumber\\
&-&({\bar{n}_{th}+{1}})({\Gamma/{2}})({{\hat{\rho}}_{at}}{\hat{\sigma}_{+}}{\hat{\sigma}_{-}}{-}{\hat{\sigma}_{-}}{{\hat{\rho}}_{at}}{\hat{\sigma}_{+}})+{H.c.},
\end{eqnarray}
and
\begin{eqnarray}\label{eq:11}
{\dot{\hat{\rho}}^{\tiny{\mbox{WRWA}}}_{at}}{(t)}&=&-{\bar{n}_{th}}({\Gamma/{2}})({{\hat{\rho}}_{at}}{\hat{\sigma}_{+}}{\hat{\sigma}_{-}}{-}2{\hat{\sigma}_{-}}{{\hat{\rho}}_{at}}{\hat{\sigma}_{-}}{-}{\hat{\sigma}_{-}}{{\hat{\rho}}_{at}}{\hat{\sigma}_{+}})\\
&-&({\bar{n}_{th}+{1}})({\Gamma/{2}})({{\hat{\rho}}_{at}}{\hat{\sigma}_{-}}{\hat{\sigma}_{+}}{-}2{\hat{\sigma}_{-}}{{\hat{\rho}}_{at}}{\hat{\sigma}_{-}}{-}{\hat{\sigma}_{+}}{{\hat{\rho}}_{at}}{\hat{\sigma}_{-}})\nonumber+{H.c.} 
\end{eqnarray}

\section{Two-level atomic dissipation in contact with squeezed vacuum reservoir}

We now consider the situation in which the two-level atom is coupled to a squeezed vacuum field reservoir. The reservoir density operator is now given by \cite{Hong}:
\begin{eqnarray}\label{eq:12}
{\hat{\rho}_{R}}={\Pi_{k}}{{\hat{S}}_{k}(\xi)} {\left |{{0}_{k}} \right \rangle} {\left \langle {0_{k}}\right |}{{\hat{S}}_{k}^{\dag}(\xi)}, 
\end{eqnarray}
where the squeezing operator reads as,
\begin{eqnarray}\label{eq:13}
{\hat{S}_{k}(\xi)}=e^{{{\xi}^{*}}{\hat{b}_{k_{0}+k}\hat{b}_{k_{0}-k}}-{{\xi}}{\hat{b}_{k_{0}+k}^{\dag}\hat{b}_{k_{0}-k}^{\dag}}},~~~~~~~~{\xi=re^{i\theta}}.
\end{eqnarray}
It can be  easily shown that for the squeezed reservoir the following relations hold,
\begin{eqnarray}\label{eq:14}
{\langle{\hat{b}_{k}}\rangle}_{ss}&=&{\langle{\hat{b}_{k}^{\dag}}\rangle}_{ss}=0,
~~~~~~~~~~~~~~~~~{\langle{\hat{b}_{k}^{\dag}}{\hat{b}_{k'}}\rangle}_{ss}={\sinh^{2}r}{\delta_{kk'}},
\nonumber\\
{\langle{\hat{b}_{k}}{\hat{b}_{k'}^{\dag}}\rangle}_{ss}&=&{\cosh^{2}r}{\delta_{kk'}},
~~~~~~~~~~~~~~~{\langle{\hat{b}_{k}^{\dag}}{\hat{b}_{k'}^{\dag}}\rangle}_{ss}=-{e^{-i\theta}}{\sinh r}~{\cosh r},~~\nonumber\\
{\langle{\hat{b}_{k}}{\hat{b}_{k'}}\rangle}_{ss}&=&-{e^{i\theta}}{\sinh r}~{\cosh r}.
\end{eqnarray}
On substituting the relations (\ref{eq:14}) into (\ref{eq:3}) one has,
\begin{eqnarray}\label{eq:15}
{{\dot{\hat{\rho}}_{at}}{(t)}}&=&{\frac{\Gamma}{2}}{{e^{i\theta}}{\sinh r}~{\cosh r}}[{\hat{\sigma}_{+}},{\hat{\sigma}_{+}}{\hat{\rho}_{at}{(t)}}]
-{\frac{\Gamma}{2}}{{e^{i\theta}}{\sinh r}~{\cosh r}}[{\hat{\sigma}_{+}},{\hat{\rho}_{at}{(t)}}{\hat{\sigma}_{+}}]\nonumber\\
&+&{\frac{\Gamma}{2}}{{e^{i\theta}}{\sinh r}~{\cosh r}}[{\hat{\sigma}_{+}},{\hat{\sigma}_{-}}{\hat{\rho}_{at}{(t)}}]
-{\frac{\Gamma}{2}}{{e^{i\theta}}{\sinh r}~{\cosh r}}[{\hat{\sigma}_{+}},{\hat{\rho}_{at}{(t)}}{\hat{\sigma}_{-}}]\nonumber\\
&-&{\frac{\Gamma}{2}}{\sinh^{2}r}[{\hat{\sigma}_{+}},{\hat{\sigma}_{+}}{\hat{\rho}_{at}{(t)}}]+{\frac{\Gamma}{2}}{\sinh^{2}r}[{\hat{\sigma}_{+}},{\hat{\rho}_{at}{(t)}}{\hat{\sigma}_{+}}]\nonumber\\
&-&{\frac{\Gamma}{2}}{\sinh^{2}r}[{\hat{\sigma}_{+}},{\hat{\sigma}_{-}}{\hat{\rho}_{at}{(t)}}]+{\frac{\Gamma}{2}}{\sinh^{2}r}[{\hat{\sigma}_{+}},{\hat{\rho}_{at}{(t)}}{\hat{\sigma}_{-}}]\nonumber\\
&-&{\frac{\Gamma}{2}}{\cosh^{2}r}[{\hat{\sigma}_{-}},{\hat{\sigma}_{-}}{\hat{\rho}_{at}{(t)}}]+{\frac{\Gamma}{2}}{\cosh^{2}r}[{\hat{\sigma}_{-}},{\hat{\rho}_{at}{(t)}}{\hat{\sigma}_{-}}]\nonumber\\
&+&{\frac{\Gamma}{2}}{{e^{i\theta}}{\sinh r}~{\cosh r}}[{\hat{\sigma}_{-}},{\hat{\sigma}_{+}}{\hat{\rho}_{at}{(t)}}]
-{\frac{\Gamma}{2}}{{e^{i\theta}}{\sinh r}~{\cosh r}}[{\hat{\sigma}_{-}},{\hat{\rho}_{at}{(t)}}{\hat{\sigma}_{+}}]\nonumber\\
&-&{\frac{\Gamma}{2}}{\cosh^{2}r}[{\hat{\sigma}_{-}},{\hat{\sigma}_{+}}{\hat{\rho}_{at}{(t)}}]+{\frac{\Gamma}{2}}{\cosh^{2}r}[{\hat{\sigma}_{-}},{\hat{\rho}_{at}{(t)}}{\hat{\sigma}_{+}}]\nonumber\\
&+&{\frac{\Gamma}{2}}{{e^{i\theta}}{\sinh r}~{\cosh r}}[{\hat{\sigma}_{-}},{\hat{\sigma}_{-}}{\hat{\rho}_{at}{(t)}}]
-{\frac{\Gamma}{2}}{{e^{i\theta}}{\sinh r}~{\cosh r}}[{\hat{\sigma}_{-}},{\hat{\rho}_{at}{(t)}}{\hat{\sigma}_{-}}]\nonumber\\
&+&H.c.
\end{eqnarray}
Following the same procedure explained above for the thermal reservoir and using the latter results for the atom affected by the squeezed vacuum reservoir, the time evolution of atomic density operator including two terms can be obtained as,  
\begin{eqnarray}\label{eq:16}
{\dot{\hat{\rho}}_{at}}{(t)}={\dot{\hat{\rho}}^{\tiny{\mbox{RWA}}}_{at}}{(t)}+{\dot{\hat{\rho}}^{\tiny{\mbox{WRWA}}}_{at}}{(t)},
\end{eqnarray}
wherein the first and second terms respectively associated with the time evolution of the atomic density operator with and without {\small{\mbox{RWA}}}, are explicitly defined as,
\begin{eqnarray}\label{eq:17}
{\dot{\hat{\rho}}^{\tiny{\mbox{RWA}}}_{at}}{(t)}&=&{-}({\Gamma/{2}}){\cosh^2 r}({\hat{\sigma}_{+}}{\hat{\sigma}_{-}}{{\hat{\rho}}_{at}}{-}{\hat{\sigma}_{-}}{{\hat{\rho}}_{at}}{\hat{\sigma}_{+}})\nonumber\\
&{-}&({\Gamma/{2}}){\sinh^2 r}({\hat{\sigma}_{-}}{\hat{\sigma}_{+}}{{\hat{\rho}}_{at}}{-}{\sigma_{+}}{{\hat{\rho}}_{at}}{\hat{\sigma}_{-}})\nonumber\\
&{-}&{\Gamma}{e^{-i\theta}}{\sinh r}~{\cosh r}({\hat{\sigma}_{-}}{{\hat{\rho}}_{at}}{\hat{\sigma}_{-}}){+}{H.c},
\end{eqnarray}
and 
\begin{eqnarray}\label{eq:18}
{\dot{\hat{\rho}}^{\tiny{\mbox{WRWA}}}_{at}}{(t)}&=&-({\Gamma/{2}}){\cosh^2 r}({{\hat{\rho}}_{at}}{\hat{\sigma}_{-}}{\hat{\sigma}_{+}}{-}2{\hat{\sigma}_{-}}{{\hat{\rho}}_{at}}{\hat{\sigma}_{-}}{-}{\hat{\sigma}_{+}}{{\hat{\rho}}_{at}}{\hat{\sigma}_{-}})\nonumber\\
&-&({\Gamma/{2}}){\sinh^2 r}({\hat{\sigma}_{+}}{\hat{\sigma}_{-}}{{\hat{\rho}}_{at}}{-}2{\hat{\sigma}_{-}}{{\hat{\rho}}_{at}}{\hat{\sigma}_{-}}{-}{\hat{\sigma}_{-}}{{\hat{\rho}}_{at}}{\hat{\sigma}_{+}})\nonumber\\
&+&({\Gamma/{2}}){e^{-i\theta}}{\sinh r}~{\cosh r}({\hat{\sigma}_{-}}{\hat{\sigma}_{+}}{{\hat{\rho}}_{at}}{-}{2}{\hat{\sigma}_{+}}{{\hat{\rho}}_{at}}{\hat{\sigma}_{-}}-{2}{\hat{\sigma}_{-}}{{\hat{\rho}}_{at}}{\hat{\sigma}_{+}}\nonumber\\
&+&{{\hat{\rho}}_{at}}{\hat{\sigma}_{+}}{\hat{\sigma}_{-}}+{\hat{\sigma}_{+}}{\hat{\sigma}_{-}}{{\hat{\rho}}_{at}}+{{\hat{\rho}}_{at}}{\hat{\sigma}_{-}}{\hat{\sigma}_{+}})\nonumber\\
&-&{\Gamma}{e^{-i\theta}}{\sinh r}~{\cosh r}({\hat{\sigma}_{+}}{{\hat{\rho}}_{at}}{\hat{\sigma}_{+}})+H.c. 
\end{eqnarray}
\section{Field dissipation in contact with thermal reservoir}
In this subsection  we want to consider the field-reservoir interaction, while again the CRTs in the field-reservoir Hamiltonian in the interaction picture are taken into account and {\mbox{RWA}} has been put aside \cite{Knight}. Accordingly, the field-reservoir Hamiltonian in the interaction picture can be written as follows,
\begin{eqnarray}\label{eq:19}
{\hat{V}(t)}={\hbar}{\sum\limits_{k}}{g_{k}}({\hat{b}^\dagger_{k}{\hat{a}}}{e^{-i(\omega-\nu_{k})
		t}}+{\hat{b}_{k}{\hat{a}^\dagger}}{e^{i(\omega-\nu_{k})t}}+{\hat{b}_{k}{\hat{a}}}{e^{-i(\omega+\nu_{k})t}}
+{\hat{b}^\dagger_{k}{\hat{a}^\dagger}}{e^{i(\omega+\nu_{k})t}}).
\end{eqnarray}
By substituting Hamiltonian (\ref{eq:19}) into Eq. (\ref{eq:1}) and tracing over the reservoir basis, the time evolution of the field density operator in contact with thermal reservoir can be straightforwardly obtained as,
\begin{eqnarray}\label{eq:20}
{{\dot{\hat{\rho}}_{f}}{(t)}}&=&-{i\sum\limits_{k}}{g_{k}}{\langle{\hat{b}_{k}}\rangle}
{[\hat{a}^{\dag}},{\hat{\rho}_{f}{(t)}]}{e^{i(\omega-\nu_{k})t}}-{i\sum\limits_{k}}{g_{k}}{\langle{\hat{b}_{k}}\rangle}{[\hat{a}},{\hat{\rho}_{f}{(t)}]}{e^{-i(\omega+\nu_{k})t}}\nonumber\\
&+&{\int\limits_{t_{i}}^{t}}{dt'}{\sum_{k}}{\sum_{k'}}{g_{k}}{g_{k'}}{\langle{\hat{b}_{k}^{\dag}}{\hat{b}_{k'}}\rangle}{[\hat{a}},{\hat{\rho}_{f}{(t')}}{\hat{a}}]{e^{-i(\omega-\nu_{k})t-i(\omega+\nu_{k'})t'}}\nonumber\\
&-&{\int\limits_{t_{i}}^{t}}{dt'}{\sum_{k}}{\sum_{k'}}{g_{k}}{g_{k'}}{\langle{\hat{b}_{k}}{\hat{b}_{k'}^{\dag}}\rangle}[{\hat{a}^{\dag}},{\hat{a}}{\hat{\rho}_{f}{(t')}}]{e^{i(\omega-\nu_{k})t-i(\omega-\nu_{k'})t'}}\nonumber\\
&+&{\int\limits_{t_{i}}^{t}}{dt'}{\sum_{k}}{\sum_{k'}}{g_{k}}{g_{k'}}{\langle{\hat{b}_{k}}{\hat{b}_{k'}^{\dag}}\rangle}[{\hat{a}^{\dag}},{\hat{\rho}_{f}{(t')}}{\hat{a}}]{e^{i(\omega-\nu_{k})t-i(\omega-\nu_{k'})t'}}\nonumber\\
&-&{\int\limits_{t_{i}}^{t}}{dt'}{\sum_{k}}{\sum_{k'}}{g_{k}}{g_{k'}}{\langle{\hat{b}_{k}}{\hat{b}_{k'}}\rangle}[{\hat{a}^{\dag}},{\hat{a}^{\dag}}{\hat{\rho}_{f}{(t')}}]{e^{i(\omega-\nu_{k})t+i(\omega-\nu_{k'})t'}}\nonumber\\
&+&{\int\limits_{t_{i}}^{t}}{dt'}{\sum_{k}}{\sum_{k'}}{g_{k}}{g_{k'}}{\langle{\hat{b}_{k}}{\hat{b}_{k'}}\rangle}[{\hat{a}^{\dag}},{\hat{\rho}_{f}{(t')}}{\hat{a}^{\dag}}]{e^{i(\omega-\nu_{k})t+i(\omega-\nu_{k'})t'}}\nonumber\\
&+&{\int\limits_{t_{i}}^{t}}{dt'}{\sum_{k}}{\sum_{k'}}{g_{k}}{g_{k'}}{\langle{\hat{b}_{k}}{\hat{b}_{k'}^{\dag}}\rangle}[{\hat{a}^{\dag}},{\hat{\rho}_{f}{(t')}}{\hat{a}^{\dag}}]{e^{i(\omega-\nu_{k})t+i(\omega+\nu_{k'})t'}}\nonumber\\
&-&{\int\limits_{t_{i}}^{t}}{dt'}{\sum_{k}}{\sum_{k'}}{g_{k}}{g_{k'}}{\langle{\hat{b}_{k}}{\hat{b}_{k'}}\rangle}[{\hat{a}^{\dag}},{\hat{a}}{\hat{\rho}_{f}{(t')}}]{e^{i(\omega-\nu_{k})t-i(\omega+\nu_{k'})t'}}\nonumber\\
&+&{\int\limits_{t_{i}}^{t}}{dt'}{\sum_{k}}{\sum_{k'}}{g_{k}}{g_{k'}}{\langle{\hat{b}_{k}}{\hat{b}_{k'}}\rangle}[{\hat{a}^{\dag}},{\hat{\rho}_{f}{(t')}}{\hat{a}}]{e^{i(\omega-\nu_{k})t-i(\omega+\nu_{k'})t'}}\nonumber\\
&-&{\int\limits_{t_{i}}^{t}}{dt'}{\sum_{k}}{\sum_{k'}}{g_{k}}{g_{k'}}{\langle{\hat{b}_{k}}{\hat{b}_{k'}^{\dag}}\rangle}[{\hat{a}},{\hat{a}}{\hat{\rho}_{f}{(t')}}]{e^{-i(\omega+\nu_{k})t-i(\omega-\nu_{k'})t'}}\nonumber
\end{eqnarray}
\begin{eqnarray}\label{Hintint}
&+&{\int\limits_{t_{i}}^{t}}{dt'}{\sum_{k}}{\sum_{k'}}{g_{k}}{g_{k'}}{\langle{\hat{b}_{k}}{\hat{b}_{k'}^{\dag}}\rangle}[{\hat{a}},{\hat{\rho}_{f}{(t')}}{\hat{a}}]{e^{-i(\omega+\nu_{k})t-i(\omega-\nu_{k'})t'}}\nonumber\\
&-&{\int\limits_{t_{i}}^{t}}{dt'}{\sum_{k}}{\sum_{k'}}{g_{k}}{g_{k'}}{\langle{\hat{b}_{k}}{\hat{b}_{k'}}\rangle}[{\hat{a}},{\hat{a}^{\dag}}{\hat{\rho}_{f}{(t')}}]{e^{-i(\omega+\nu_{k})t+i(\omega-\nu_{k'})t'}}\nonumber\\
&+&{\int\limits_{t_{i}}^{t}}{dt'}{\sum_{k}}{\sum_{k'}}{g_{k}}{g_{k'}}{\langle{\hat{b}_{k}}{\hat{b}_{k'}}\rangle}[{\hat{a}},{\hat{\rho}_{f}{(t')}}{\hat{a}^{\dag}}]{e^{-i(\omega+\nu_{k})t+i(\omega-\nu_{k'})t'}}\nonumber\\
&-&{\int\limits_{t_{i}}^{t}}{dt'}{\sum_{k}}{\sum_{k'}}{g_{k}}{g_{k'}}{\langle{\hat{b}_{k}}{\hat{b}_{k'}^{\dag}}\rangle}[{\hat{a}},{\hat{a}^{\dag}}{\hat{\rho}_{f}{(t')}}]{e^{-i(\omega+\nu_{k})t+i(\omega+\nu_{k'})t'}}\nonumber\\
&+&{\int\limits_{t_{i}}^{t}}{dt'}{\sum_{k}}{\sum_{k'}}{g_{k}}{g_{k'}}{\langle{\hat{b}_{k}}{\hat{b}_{k'}^{\dag}}\rangle}[{\hat{a}},{\hat{\rho}_{f}{(t')}}{\hat{a}^{\dag}}]{e^{-i(\omega+\nu_{k})t+i(\omega+\nu_{k'})t'}}\nonumber\\
&-&{\int\limits_{t_{i}}^{t}}{dt'}{\sum_{k}}{\sum_{k'}}{g_{k}}{g_{k'}}{\langle{\hat{b}_{k}}{\hat{b}_{k'}}\rangle}[{\hat{a}},{\hat{a}}{\hat{\rho}_{f}{(t')}}]{e^{-i(\omega+\nu_{k})t-i(\omega+\nu_{k'})t'}}\nonumber\\
&+&{\int\limits_{t_{i}}^{t}}{dt'}{\sum_{k}}{\sum_{k'}}{g_{k}}{g_{k'}}{\langle{\hat{b}_{k}}{\hat{b}_{k'}}\rangle}[{\hat{a}},{\hat{\rho}_{f}{(t')}}{\hat{a}}]{e^{-i(\omega+\nu_{k})t-i(\omega+\nu_{k'})t'}}+H.c.
\end{eqnarray}
Now, following the same procedure led us to the atomic decay rate in Eq. (\ref{eq:8}), for the explicit form of of the decay rate of the filed arrives one to, 
\begin{eqnarray}\label{eq:21}
\frac{2V}{{(2\pi )}^{3}}{\int\limits_{t_{i}}^{t}}{dt'}{\int\limits_{0}^{2\pi}}{d\phi}{\int\limits_{0}^{\pi}}{d\theta}~{\sin{\theta}}{\int\limits_{0}^{\infty}}{dk}~{k^{2}}=\frac{V{{\nu}^{2}}}{{\pi }^{2}{c }^{3}}{d\nu}.
\end{eqnarray}
Now,since  the  the number of modes with frequencies in the range {$\nu$} to {$\nu+d\nu$} is reads as, 
\begin{eqnarray}\label{eq:22}
D({\nu}){d{\nu}}=\frac{V{{\nu}^{2}}}{{\pi }^{2}{c }^{3}}{d\nu},
\end{eqnarray}
with {$D({\nu})$} as the mode density, we finally obtain, 
\begin{eqnarray}\label{eq:23}
{\int\limits_{t_{i}}^{t}}{dt'}{\sum_{k}}{g_{k}^{2}}{e^{-i(\omega+\nu_{k})t+i(\omega+\nu_{k})t'}}={2\pi}{g_{\frac{\nu}{c}}^{2}}D({\nu})=C,
\end{eqnarray}
where  C now plays the role of field damping. \\
Now, via substituting the expectation values introduced in (\ref{eq:4}) into (\ref{eq:20}), we may obtain, 
\begin{eqnarray}\label{eq:24}
{{\dot{\hat{\rho}}_{f}}{(t)}}&=&{-}{\frac{C}{2}}{\bar{n}_{th}}{[\hat{a}},{\hat{a}^{\dag}}{\hat{\rho}_{f}{(t)}}]+{\frac{C}{2}}{\bar{n}_{th}}{[\hat{a}},{\hat{\rho}_{f}{(t)}{\hat{a}^{\dag}}]}-{\frac{C}{2}}{\bar{n}_{th}}{[\hat{a}},{\hat{a}}{\hat{\rho}_{f}{(t)}}]+{\frac{C}{2}}{\bar{n}_{th}}{[\hat{a}},{\hat{\rho}_{f}{(t)}}{\hat{a}}]\nonumber\\
&-&{\frac{C}{2}}{\bar{n}_{th}}[{\hat{a}^{\dag}},{\hat{a}^{\dag}}{\hat{\rho}_{f}{(t)}}]
+{\frac{C}{2}}{\bar{n}_{th}}[{\hat{a}^{\dag}},{\hat{\rho}_{at}{(t)}}{\hat{a}^{\dag}}]+{\frac{C}{2}}{\bar{n}_{th}}[{\hat{a}^{\dag}},{\hat{\rho}_{f}{(t)}}{\hat{a}^{\dag}}]
-{\frac{C}{2}}{\bar{n}_{th}}[{\hat{a}^{\dag}},{\hat{a}}{\hat{\rho}_{f}{(t)}}]
\nonumber\\
&+&{\frac{C}{2}}{\bar{n}_{th}}[{\hat{a}^{\dag}},{\hat{\rho}_{f}{(t)}}{\hat{a}}]
+H.c.
\end{eqnarray}
Finally, we achieve the time evolution of field density operator for thermal  reservoir as follows,
\begin{eqnarray}\label{eq:25}
{\dot{\hat{\rho}}_{f}}{(t)}={\dot{\hat{\rho}}^{\tiny{\mbox{RWA}}}_{f}}{(t)}+{\dot{\hat{\rho}}^{\tiny{\mbox{WRWA}}}_{f}}{(t)},
\end{eqnarray}
wherein the first and second terms are respectively associated with the time evolution of the field density operator with and without {\small{\mbox{RWA}}}, have been obtained as, 
\begin{eqnarray}\label{eq:26}
{\dot{\hat{\rho}}^{\tiny{\mbox{RWA}}}_{f}}{(t)}&=&-({C/{2}}){n_{th}}({\hat{a}}{\hat{a}^\dagger}{{\hat{\rho}_{f}}}{-}2{\hat{a}^\dagger}{{\hat{\rho}_{f}}}{\hat{a}}+{{\hat{\rho}_{f}}}{\hat{a}}{\hat{a}^\dagger})\nonumber\\
&-&({C/{2}})({n_{th}+1})({\hat{a}^\dagger}{\hat{a}}{{\hat{\rho}_{f}}}{-}2{\hat{a}}{{\hat{\rho}_{f}}}{\hat{a}^\dagger}{+}{{\hat{\rho}_{f}}}{\hat{a}^\dagger}{\hat{a}}),
\end{eqnarray}
and
\begin{eqnarray}\label{eq:27}
{\dot{\hat{\rho}}^{\tiny{\mbox{WRWA}}}_{f}}{(t)}&=&-({C/{2}}){n_{th}}({-}{\hat{a}}{{\hat{\rho}_{f}}}{\hat{a}^\dagger}{+}{{\hat{\rho}_{f}}}{\hat{a}^\dagger}{\hat{a}}+{\hat{a}}{\hat{a}}{{\rho}_{f}}{-}2{\hat{a}}{{\hat{\rho}_{f}}}{\hat{a}}+{{\hat{\rho}_{f}}}{\hat{a}}{\hat{a}})\nonumber\\
&-&({C/{2}})({n_{th}+1})({-}{\hat{a}^\dagger}{{\hat{\rho}_{f}}}{\hat{a}}{+}{{\hat{\rho}_{f}}}{\hat{a}}{\hat{a}^\dagger}{+}{\hat{a}^\dagger}{\hat{a}^\dagger}{{\hat{\rho}_{f}}}{-}2{\hat{a}^\dagger}{{\hat{\rho}}}{\hat{a}^\dagger}+{\hat{a}}{\hat{a}}{{\hat{\rho}_{f}}})+{H.c}.
\end{eqnarray}
\section{Field dissipation in contact with squeezed vacuum reservoir}

In the following we consider the field-reservoir interaction system while the  reservoir constitutes from squeezed vacuum states  \cite{Scully} and CRTs are taken into account, i.e., without {\mbox{RWA}} (see Eq. (\ref{eq:19})).
Now, by substituting  (\ref{eq:19}) into (\ref{eq:1}) and  via tracing over the reservoir modes, the  time evolution of the field density operator in contact with a squeezed reservoir can be obtained as,

\begin{eqnarray}\label{eq:28}
{{\dot{\hat{\rho}}_{f}}{(t)}}&=&-{i\sum\limits_{k}}{g_{k}}{\langle{\hat{b}_{k}}\rangle}_{ss}
{[\hat{a}^{\dag}},{\hat{\rho}_{f}{(t)}]}{e^{i(\omega-\nu_{k})t}}-{i\sum\limits_{k}}{g_{k}}{\langle{\hat{b}_{k}}\rangle}_{ss}{[\hat{a}},{\hat{\rho}_{f}{(t)}]}{e^{-i(\omega+\nu_{k})t}}\nonumber\\
&-&{\int\limits_{t_{i}}^{t}}{dt'}{\sum_{k}}{\sum_{k'}}{g_{k}}{g_{k'}}{\langle{\hat{b}_{k}}{\hat{b}_{k'}^{\dag}}\rangle}_{ss}[{\hat{a}^{\dag}},{\hat{a}}{\hat{\rho}_{f}{(t')}}]{e^{i(\omega-\nu_{k})t-i(\omega-\nu_{k'})t'}}\nonumber\\
&+&{\int\limits_{t_{i}}^{t}}{dt'}{\sum_{k}}{\sum_{k'}}{g_{k}}{g_{k'}}{\langle{\hat{b}_{k}}{\hat{b}_{k'}^{\dag}}\rangle}_{ss}[{\hat{a}^{\dag}},{\hat{\rho}_{f}{(t')}}{\hat{a}}]{e^{i(\omega-\nu_{k})t-i(\omega-\nu_{k'})t'}}\nonumber\\
&-&{\int\limits_{t_{i}}^{t}}{dt'}{\sum_{k}}{\sum_{k'}}{g_{k}}{g_{k'}}{\langle{\hat{b}_{k}}{\hat{b}_{k'}}\rangle}_{ss}[{\hat{a}^{\dag}},{\hat{a}^{\dag}}{\hat{\rho}_{f}{(t')}}]{e^{i(\omega-\nu_{k})t+i(\omega-\nu_{k'})t'}}\nonumber\\
&+&{\int\limits_{t_{i}}^{t}}{dt'}{\sum_{k}}{\sum_{k'}}{g_{k}}{g_{k'}}{\langle{\hat{b}_{k}}{\hat{b}_{k'}}\rangle}_{ss}[{\hat{a}^{\dag}},{\hat{\rho}_{f}{(t')}}{\hat{a}^{\dag}}]{e^{i(\omega-\nu_{k})t+i(\omega-\nu_{k'})t'}}\nonumber\\
&-&{\int\limits_{t_{i}}^{t}}{dt'}{\sum_{k}}{\sum_{k'}}{g_{k}}{g_{k'}}{\langle{\hat{b}_{k}}{\hat{b}_{k'}^{\dag}}\rangle}_{ss}[{\hat{a}^{\dag}},{\hat{a}^{\dag}}{\hat{\rho}_{f}{(t')}}]{e^{i(\omega-\nu_{k})t+i(\omega+\nu_{k'})t'}}\nonumber\\
&+&{\int\limits_{t_{i}}^{t}}{dt'}{\sum_{k}}{\sum_{k'}}{g_{k}}{g_{k'}}{\langle{\hat{b}_{k}}{\hat{b}_{k'}^{\dag}}\rangle}_{ss}[{\hat{a}^{\dag}},{\hat{\rho}_{f}{(t')}}{\hat{a}^{\dag}}]{e^{i(\omega-\nu_{k})t+i(\omega+\nu_{k'})t'}}\nonumber\\
&-&{\int\limits_{t_{i}}^{t}}{dt'}{\sum_{k}}{\sum_{k'}}{g_{k}}{g_{k'}}{\langle{\hat{b}_{k}}{\hat{b}_{k'}}\rangle}_{ss}[{\hat{a}^{\dag}},{\hat{a}}{\hat{\rho}_{f}{(t')}}]{e^{i(\omega-\nu_{k})t-i(\omega+\nu_{k'})t'}}\nonumber\\
&+&{\int\limits_{t_{i}}^{t}}{dt'}{\sum_{k}}{\sum_{k'}}{g_{k}}{g_{k'}}{\langle{\hat{b}_{k}}{\hat{b}_{k'}}\rangle}_{ss}[{\hat{a}^{\dag}},{\hat{\rho}_{f}{(t')}}{\hat{a}}]{e^{i(\omega-\nu_{k})t-i(\omega+\nu_{k'})t'}}\nonumber\\
&+&{\int\limits_{t_{i}}^{t}}{dt'}{\sum_{k}}{\sum_{k'}}{g_{k}}{g_{k'}}{\langle{\hat{b}_{k}^{\dag}}{\hat{b}_{k'}^{\dag}}\rangle}_{ss}[{\hat{a}^{\dag}},{\hat{\rho}_{f}{(t')}}{\hat{a}}]{e^{i(\omega+\nu_{k})t-i(\omega-\nu_{k'})t'}}\nonumber\\
&-&{\int\limits_{t_{i}}^{t}}{dt'}{\sum_{k}}{\sum_{k'}}{g_{k}}{g_{k'}}{\langle{\hat{b}_{k}}{\hat{b}_{k'}^{\dag}}\rangle}_{ss}[{\hat{a}},{\hat{a}}{\hat{\rho}_{f}{(t')}}]{e^{-i(\omega+\nu_{k})t-i(\omega-\nu_{k'})t'}}\nonumber\\
&+&{\int\limits_{t_{i}}^{t}}{dt'}{\sum_{k}}{\sum_{k'}}{g_{k}}{g_{k'}}{\langle{\hat{b}_{k}}{\hat{b}_{k'}^{\dag}}\rangle}_{ss}[{\hat{a}},{\hat{\rho}_{f}{(t')}}{\hat{a}}]{e^{-i(\omega+\nu_{k})t-i(\omega-\nu_{k'})t'}}\nonumber\\
&-&{\int\limits_{t_{i}}^{t}}{dt'}{\sum_{k}}{\sum_{k'}}{g_{k}}{g_{k'}}{\langle{\hat{b}_{k}}{\hat{b}_{k'}}\rangle}_{ss}[{\hat{a}},{\hat{a}^{\dag}}{\hat{\rho}_{f}{(t')}}]{e^{-i(\omega+\nu_{k})t+i(\omega-\nu_{k'})t'}}\nonumber\\
&+&{\int\limits_{t_{i}}^{t}}{dt'}{\sum_{k}}{\sum_{k'}}{g_{k}}{g_{k'}}{\langle{\hat{b}_{k}}{\hat{b}_{k'}}\rangle}_{ss}[{\hat{a}},{\hat{\rho}_{f}{(t')}}{\hat{a}^{\dag}}]{e^{-i(\omega+\nu_{k})t+i(\omega-\nu_{k'})t'}}\nonumber
\end{eqnarray}
\begin{eqnarray}\label{Hintint}
&-&{\int\limits_{t_{i}}^{t}}{dt'}{\sum_{k}}{\sum_{k'}}{g_{k}}{g_{k'}}{\langle{\hat{b}_{k}}{\hat{b}_{k'}^{\dag}}\rangle}_{ss}[{\hat{a}},{\hat{a}^{\dag}}{\hat{\rho}_{f}{(t')}}]{e^{-i(\omega+\nu_{k})t+i(\omega+\nu_{k'})t'}}\nonumber\\
&+&{\int\limits_{t_{i}}^{t}}{dt'}{\sum_{k}}{\sum_{k'}}{g_{k}}{g_{k'}}{\langle{\hat{b}_{k}}{\hat{b}_{k'}^{\dag}}\rangle}_{ss}[{\hat{a}},{\hat{\rho}_{f}{(t')}}{\hat{a}^{\dag}}]{e^{-i(\omega+\nu_{k})t+i(\omega+\nu_{k'})t'}}\nonumber\\
&-&{\int\limits_{t_{i}}^{t}}{dt'}{\sum_{k}}{\sum_{k'}}{g_{k}}{g_{k'}}{\langle{\hat{b}_{k}}{\hat{b}_{k'}}\rangle}_{ss}[{\hat{a}},{\hat{a}}{\hat{\rho}_{f}{(t')}}]{e^{-i(\omega+\nu_{k})t-i(\omega+\nu_{k'})t'}}\nonumber\\
&+&{\int\limits_{t_{i}}^{t}}{dt'}{\sum_{k}}{\sum_{k'}}{g_{k}}{g_{k'}}{\langle{\hat{b}_{k}}{\hat{b}_{k'}}\rangle}_{ss}[{\hat{a}},{\hat{\rho}_{f}{(t')}}{\hat{a}}]{e^{-i(\omega+\nu_{k})t-i(\omega+\nu_{k'})t'}}.
\end{eqnarray}

On substituting the expectation values in (\ref{eq:29}) from (\ref{eq:14}), we finally arrive at the time evolution of field density operator for squeezed reservoir as,

\begin{eqnarray}\label{eq:29}
{{\dot{\hat{\rho}}_{f}}{(t)}}&=&{\frac{C}{2}}{{e^{-i\theta}}{\sinh r}~{\cosh r}}{[{\hat{a}},{\hat{a}}{\hat{\rho}_{f}{(t)}}]}-{\frac{C}{2}}{{e^{-i\theta}}{\sinh r}~{\cosh r}}[{\hat{a}},{\hat{\rho}_{f}{(t)}}{\hat{a}}]\nonumber\\
&-&{\frac{C}{2}}{\sinh^{2}r}[{\hat{a}},{\hat{a}^{\dag}}{\hat{\rho}_{f}{(t)}}]+{\frac{C}{2}}{\sinh^{2}r}[{\hat{a}},{\hat{\rho}_{f}{(t)}}{\hat{a}^{\dag}}]\nonumber\\
&+&{\frac{C}{2}}{{e^{-i\theta}}{\sinh r}~{\cosh r}}[{\hat{a}},{\hat{a}^{\dag}}{\hat{\rho}_{f}{(t)}}]-{\frac{C}{2}}{{e^{-i\theta}}{\sinh r}~{\cosh r}}[{\hat{a}},{\hat{\rho}_{f}{(t)}}{\hat{{a}^{\dag}}}]\nonumber\\
&-&{\frac{C}{2}}{\sinh^{2}r}[{\hat{a}},{\hat{a}}{\hat{\rho}_{f}{(t)}}]+{\frac{C}{2}}{\sinh^{2}r}[{\hat{a}},{\hat{\rho}_{f}{(t)}}{\hat{a}}]-{\frac{C}{2}}{\cosh^{2}r}[{\hat{{a}^{\dag}}},{\hat{a}}{\hat{\rho}_{f}{(t)}}]\nonumber\\
&+&{\frac{C}{2}}{\cosh^{2}r}[{\hat{{a}^{\dag}}},{\hat{\rho}_{f}{(t)}}{\hat{a}}]-{\frac{C}{2}}{\cosh^{2}r}[{\hat{{a}^{\dag}}},{\hat{{a}^{\dag}}}{\hat{\rho}_{f}{(t)}}]+{\frac{C}{2}}{\cosh^{2}r}[{\hat{{a}^{\dag}}},{\hat{\rho}_{f}{(t)}}{\hat{{a}^{\dag}}}]\nonumber\\
&-&{\frac{C}{2}}{{e^{-i\theta}}{\sinh r}~{\cosh r}}[{\hat{{a}^{\dag}}},{\hat{\rho}_{f}{(t)}}{\hat{a}}]+{\frac{C}{2}}{{e^{-i\theta}}{\sinh r}~{\cosh r}}[{\hat{{a}^{\dag}}},{\hat{{a}^{\dag}}}{\hat{\rho}_{f}{(t)}}]\nonumber\\
&-&{\frac{C}{2}}{{e^{-i\theta}}{\sinh r}~{\cosh r}}[{\hat{{a}^{\dag}}},{\hat{\rho}_{f}{(t)}}{\hat{{a}^{\dag}}}]-{\frac{C}{2}}{{e^{i\theta}}{\sinh r}~{\cosh r}}[{\hat{{a}}},{\hat{\rho}_{f}{(t)}}{\hat{{a}^{\dag}}}]\nonumber\\
&+&{\frac{C}{2}}{{e^{i\theta}}{\sinh r}~{\cosh r}}[{\hat{{a}}},{\hat{{a}}}{\hat{\rho}_{f}{(t)}}].
\end{eqnarray}

Now by calculating the commutations relations in the above equation and simplifying its right hand side, one  obtains the time evolution of the field density operator including two terms as follows,
\begin{eqnarray}\label{eq:30}
{\dot{\hat{\rho}}_{f}}{(t)}={\dot{\hat{\rho}}^{\tiny{\mbox{RWA}}}_{f}}{(t)}+{\dot{\hat{\rho}}^{\tiny{\mbox{WRWA}}}_{f}}{(t)},
\end{eqnarray}
wherein the first and second terms are respectively associated with the time evolution of the field density operator with and without {\small{\mbox{RWA}}},

\begin{eqnarray}\label{eq:31}
{\dot{\hat{\rho}}^{\tiny{\mbox{RWA}}}_{f}}{(t)}&=&-({C/{2}}){\sinh^2r}({\hat{a}}{\hat{a}^\dagger}{{\hat{\rho}}_{f}}{-}{\hat{a}^\dagger}{{\hat{\rho}}_{f}}{\hat{a}})-({C/{2}}){\cosh^2r}({\hat{a}^\dagger}{\hat{a}}{{\hat{\rho}}_{f}}{-}{\hat{a}}{{\hat{\rho}}_{f}}{\hat{a}^\dagger})\nonumber\\
&+&({C/{2}}){\sinh r}{\cosh r}{e^{-i\theta}}({\hat{a}}{\hat{a}}{{\hat{\rho}}_{f}}{-}2{\hat{a}}{{\hat{\rho}}_{f}}{\hat{a}}+{{\hat{\rho}}_{f}}{\hat{a}}{\hat{a}})+H.c.
\end{eqnarray}
and 
\begin{eqnarray}\label{eq:32}
{\dot{\hat{\rho}}^{\tiny{\mbox{WRWA}}}_{f}}{(t)}&=&-({C/{2}}){\sinh^2r}({-}{\hat{a}}{{\hat{\rho}}_{f}}{\hat{a}^\dagger}+{{\hat{\rho}}_{f}}{\hat{a}^\dagger}{\hat{a}}+{\hat{a}}{\hat{a}}{{\hat{\rho}}_{f}}{-}2{\hat{a}}{{\hat{\rho}}_{f}}{\hat{a}}+{{\hat{\rho}}_{f}}{\hat{a}}{\hat{a}})\nonumber\\
&-&({C/{2}}){\cosh^2r}({-}{\hat{a}^\dagger}{{\hat{\rho}}_{f}}{\hat{a}}+{{\hat{\rho}}_{f}}{\hat{a}}{\hat{a}^\dagger}+{\hat{a}^\dagger}{\hat{a}^\dagger}{{\hat{\rho}}_{f}}{-}2{\hat{a}^\dagger}{{\hat{\rho}}_{f}}{\hat{a}^\dagger}+{\hat{a}}{\hat{a}}{{\hat{\rho}}_{f}})\nonumber\\
&+&({C/{2}}){{\sinh r}}{\cosh r}{e^{-i\theta}}({\hat{a}}{\hat{a}^\dagger}{{\hat{\rho}}_{f}}{+}{{\hat{\rho}}_{f}}{\hat{a}}{\hat{a}^\dagger}{-}2{\hat{a}}{{\hat{\rho}}_{f}}{\hat{a}^\dagger}{-}2{\hat{a}^\dagger}{{\hat{\rho}}_{f}}{\hat{a}}\nonumber\\
&+&{{\hat{\rho}}_{f}}{\hat{a}^\dagger}{\hat{a}}+{\hat{a}^\dagger}{\hat{a}}{{\hat{\rho}}_{f}}+{\hat{a}^\dagger}{\hat{a}^\dagger}{{\hat{\rho}}_{f}}{-}2{\hat{a}^\dagger}{{\hat{\rho}}_{f}}{\hat{a}^\dagger}+{{\hat{\rho}}_{f}}{\hat{a}^\dagger}{\hat{a}^\dagger})+{H.c}.
\end{eqnarray}


\section{Some applications of the formalism: Some physical quantities}

The results obtained in relations (\ref{eq:11}), (\ref{eq:18}), (\ref{eq:27}) and (\ref{eq:32}) clearly show that the descriptive terms of {$\dot{\hat{\rho}}^{\tiny{\mbox{WRWA}}}_{at(f)}$} have been  arisen from the CRTs terms of the Hamiltonian of the considered systems. As an  application of the approach presented above, the equations of motion for the expectation values of the hermitian atomic operators {$\hat{\sigma}_{x}$}, {$\hat{\sigma}_{y}$} and {$\hat{\sigma}_{z}$} may be obtained in the continuation for both squeezed and thermal reservoirs \cite{Zhang3}.
Considering the two mentioned reservoirs in above we arrive at,

\begin{eqnarray}\label{eq:33}
{\langle\hat{\dot\sigma}_{x}\rangle_{ss}}&=&{\langle\hat{\dot\sigma}_{x}\rangle_{th}}=0,\nonumber\\
{\langle\hat{\dot\sigma}_{y}\rangle_{ss}}&=&{-}{2}{\Gamma}{\cosh^2{r}}{\langle\hat{\sigma}_{y}\rangle_{ss}}
{-}{2}{\Gamma}{\langle\hat{\sigma}_{y}\rangle_{ss}}-{4}{\Gamma}{\cos{\theta}}~{\sinh{r}\cosh{r}}{\langle\hat{\sigma}_{y}\rangle_{ss}},\nonumber\\
{\langle\hat{\dot\sigma}_{y}\rangle_{th}}&=&{-}{2}{\Gamma}{(2n_{th}+1)}{\langle\hat{\sigma}_{y}\rangle_{th}},\nonumber\\
{\langle\hat{\dot\sigma}_{z}\rangle_{ss}}&=&{-\Gamma}{\cosh^2{r}}{\langle\hat{\sigma}_{z}\rangle_{ss}}{-}{\Gamma}{\sinh^2{r}}{\langle\hat{\sigma}_{z}\rangle_{ss}}+{2}{\Gamma}{\cos{\theta}}~{\sinh{r}\cosh{r}}{\langle\hat{\sigma}_{z}\rangle_{ss}},\nonumber\\
{\langle\hat{\dot\sigma}_{z}\rangle_{th}}&=&{-\Gamma}{(2n_{th}+1)}{\langle\hat{\sigma}_{z}\rangle_{th}}.
\end{eqnarray}
In Fig. 2 we plotted $\langle\hat{\sigma}_{y}\rangle_{ss}$ in terms of $\Gamma t$ for fixed value of $\theta=0$ different values of $r$ with and without RWA.
As is seen, in all cases the damping causes smoothly reduction of {$\langle\hat{\sigma}_{y}\rangle$} as time is passed, in such a way that it tends to zero for large {$\Gamma$}{$t$}. 
In more detail, we observe that in the absence (presence) of RWA by increasing the squeezing parameter {${r}$}, the rate of reduction {$\langle\hat{\sigma}_{y}\rangle$} increases (decreases).       
\begin{figure}[!h]
	\includegraphics[width=6.3cm]{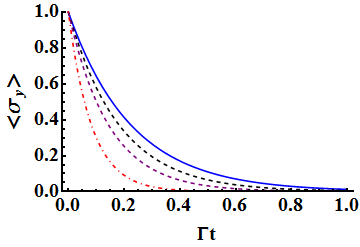}
	\includegraphics[width=6.3cm]{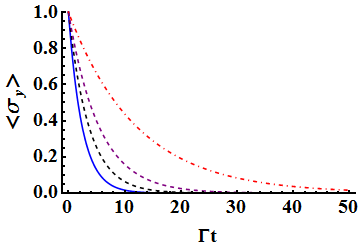}
	\caption{Left (right) figure: {{$ \langle\hat\sigma_{y}\rangle_{ss}$} in terms of the scaled time {$\Gamma$}{$t$} without (with) {\small{\mbox{RWA}}}. Solid blue line (r=0.1), dotted black line (r=0.3), dashed purple line (r=0.5) and dot-dashed red line (r=0.9), for {$\theta=0$}}.}
\end{figure}
~\\
~\\
In Fig. 3 we plotted $\langle\hat{\sigma}_{y}\rangle_{ss}$ vs $\theta$ for fixed value of $\Gamma t=1$ and different values of $r$ with and without RWA.
As is seen from this figure this quantity periodically oscillates, while the domain of oscillation increases (decreases) with increasing (decreasing) $r$ for without (with) RWA. Also, in without RWA, there exist some time intervals in which this quantity becomes zero, while in RWA this quantity never gets zero value.       

\begin{figure}[!h]
	\includegraphics[width=6.3cm]{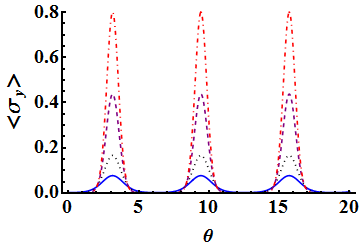}
	\includegraphics[width=6.3cm]{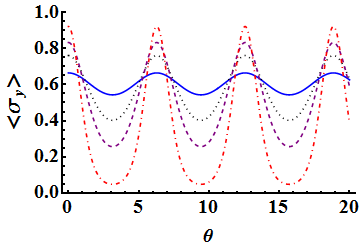}
	\caption{Left (right) figure: {$\langle\hat\sigma_{y}\rangle_{ss}$} in terms of {$\theta$} without (with) {\small{\mbox{RWA}}} for {{$\Gamma$}{$t$}=1}. Solid blue line (r=0.1), dotted black line (r=0.3), dashed purple line (r=0.5) and dot-dashed red line (r=0.9).}
\end{figure}
In Fig. 4 we plotted $\langle\hat{\sigma}_{y}\rangle_{th}$ in terms of $\Gamma t$ for different values of {$\Gamma$}{$t$} without and with RWA. As is seen, in all cases the damping causes smoothly reduction of {$\langle\hat{\sigma}_{y}\rangle$} as time is passed, in such a way that it tends to zero for large {$\Gamma$}{$t$}. 
In more detail, we observe that in the absence (presence) of RWA by increasing {$n_{th}$}, the rate of reduction {$\langle\hat{\sigma}_{y}\rangle$} increases (increases).   
\begin{figure}[!h]
	\includegraphics[width=6.3cm]{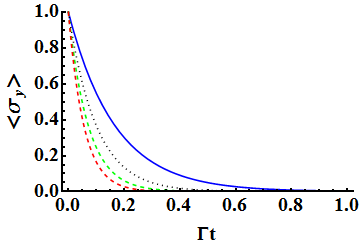}
	\includegraphics[width=6.3cm]{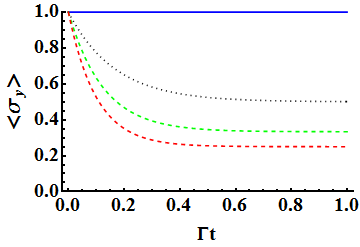}
	\caption{Left (right) figure: {$ \langle\hat\sigma_{y}\rangle_{th}$} in terms of {$n_{th}$} without (with) {\small{\mbox{RWA}}} in which, solid blue line ({$n_{th}$}=1), dotted black line ({$n_{th}$}=2), dashed purple line ({$n_{th}$}=3) and dot-dashed red line ({$n_{th}$}=4).}
\end{figure}
\newpage
In Fig. 5 we plotted $\langle\hat{\sigma}_{y}\rangle_{th}$ in terms of {$n_{th}$} for different values of {$\Gamma$}{$t$} without and with RWA.
In more detail, we observe that in the absence (presence) of RWA by increasing {$n_{th}$}, the rate of reduction {$\langle\hat{\sigma}_{y}\rangle$} increases (decreases).     

\begin{figure}[!h]
	\includegraphics[width=6.3cm]{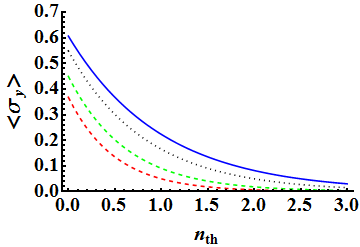}
	\includegraphics[width=6.3cm]{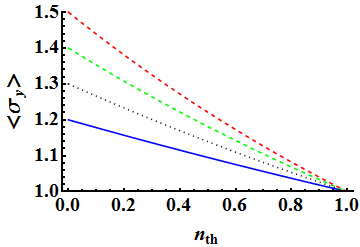}
	\caption{Left (right) figure: {$ \langle\hat\sigma_{y}\rangle_{th}$} in terms of {$n_{th}$} without (with) {\small{\mbox{RWA}}} in which, solid blue line
		({$\Gamma$}{$t$}=0.1), dotted black line ({$\Gamma$}{$t$}=0.2), dashed purple line ({$\Gamma$}{$t$}=0.3) and dot-dashed red line ({$\Gamma$}{$t$}=0.4).} 
\end{figure}
In Fig. 6 one can see that generally {$ \langle\hat\sigma_{z}\rangle_{ss}$} decreases gradually with {$\Gamma$}{$t$}. However, it should be mentioned that for a constant value of {$\Gamma$}{$t$}, this quantity can be improved (degraded) by decreasing (increasing) the squeezing parameter, specially at $\theta=0$, as shown in left without RWA and right RWA panels, respectively.        
\begin{figure}[!h]
	\includegraphics[width=6.3cm]{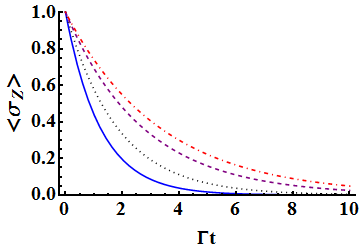}
	\includegraphics[width=6.3cm]{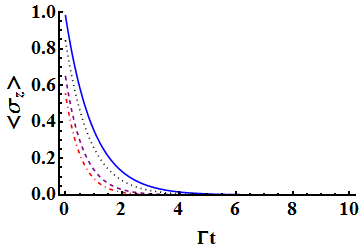}
	\caption{Left (right) figure: {{$ \langle\hat\sigma_{z}\rangle_{ss}$} in terms of the scaled time {$\Gamma$}{$t$} without (with) {\small{\mbox{RWA}}}. Solid blue line
			({$r$}=0.1), dotted black line ({$r$}=0.3), dashed purple line ({$r$}=0.5) and dot-dashed ({$r$}=0.6)} for {$\theta=0$}.}
\end{figure}
\newpage
In Fig. 7 we plotted {{$ \langle\hat\sigma_{z}\rangle_{ss}$} in terms of {$r$} without (with) {\small{\mbox{RWA}}}. We observe in without {\small{\mbox{RWA}}} that by passing time and increasing {$r$} damping is increased. But, we observe in RWA regime that the damping decreases by increasing the amount of {$\Gamma$}{$t$} and {$r$}.   
	\begin{figure}[!h]
		\includegraphics[width=6.3cm]{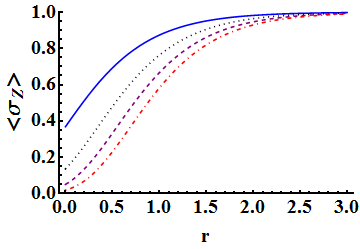}
		\includegraphics[width=6.3cm]{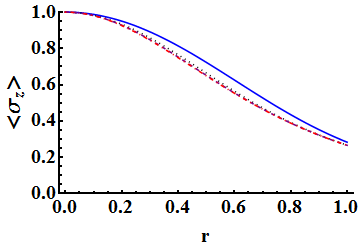}
		\caption{Left (right) figure: {{$ \langle\hat\sigma_{z}\rangle_{ss}$} in terms of {$r$} without (with) {\small{\mbox{RWA}}}. Solid blue line
				({$\Gamma$}{$t$}=1), dotted black line ({$\Gamma$}{$t$}=2), dashed purple line ({$\Gamma$}{$t$}=3) and dot-dashed ({$\Gamma$}{$t$}=4)} for {$\theta=0$}.}
	\end{figure}\\
	In Fig. 8 we plotted $\langle\hat{\sigma}_{z}\rangle_{th}$ in terms of $\Gamma t$ for different values of {$n_{th}$} with and without RWA.
	As is seen, in all cases the damping causes smoothly reduction of {$\langle\hat{\sigma}_{z}\rangle$} as time is passed, in such a way that it tends to nonzero (zero) for large {$\Gamma$}{$t$}. 
	In more detail, we observe that in the absence (presence) of RWA by increasing {$n_{th}$}, the rate of reduction {$\langle\hat{\sigma}_{z}\rangle$} increases.       		
	\begin{figure}[!h]
		\includegraphics[width=6.3cm]{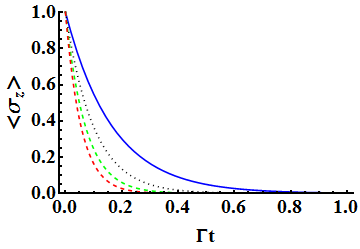}
		\includegraphics[width=6.3cm]{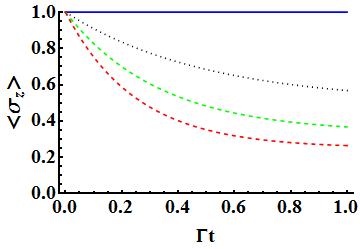}
		\caption{Left (right) figure: {{$ \langle\hat\sigma_{z}\rangle_{th}$} in terms of {$\Gamma$}{$t$} without (with) {\small{\mbox{RWA}}}. Solid blue line ({$n_{th}$}=1), dotted black line ({$n_{th}$}=2), dashed purple line ({$n_{th}$}=3) and dot-dashed ({$n_{th}$}=4)}.}
	\end{figure}
	
	
	In Fig. 9 we plotted $\langle\hat{\sigma}_{z}\rangle_{th}$ in terms of {$n_{th}$} for different values of rescaled time with and without RWA.
	As is seen, in all cases the damping causes smoothly reduction of {$\langle\hat{\sigma}_{z}\rangle$} as time is passed, in such a way that it tends to zero for large {$n_{th}$}. 
	In more detail, we observe that in the absence (presence) of RWA by increasing time, the rate of reduction {$\langle\hat{\sigma}_{z}\rangle$} increases (decreases).       
	
	\begin{figure}[!h]
		\includegraphics[width=6.3cm]{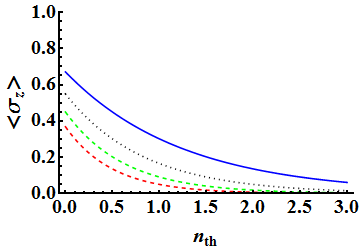}
		\includegraphics[width=6.3cm]{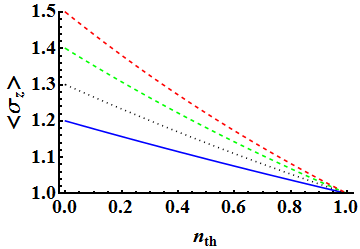}
		\caption{Left (right) figure: {$ \langle\hat\sigma_{z}\rangle_{th}$} in terms of {$n_{th}$} without (with) {\small{\mbox{RWA}}}. Solid blue line ({$\Gamma$}{$t$}=0.2), dotted black line ({$\Gamma$}{$t$}=0.3), dashed purple line ({$\Gamma$}{$t$}=0.4) and dot-dashed ({$\Gamma$}{$t$}=0.5).}
	\end{figure}
	
	In Fig. 10 we plotted {$ \langle\hat\sigma_{z}\rangle_{th }$} in terms of the scaled time {$\Gamma t$} and {$r$} without (with) {\small{\mbox{RWA}}}. We observe in without {\small{\mbox{RWA}}} by increasing time and {$r$} damping is increased. But we observe that with {\small{\mbox{RWA}}} by passing time and increasing {$r$} damping is decreased.   
	\begin{figure}[!h]
		\includegraphics[width=6.3cm]{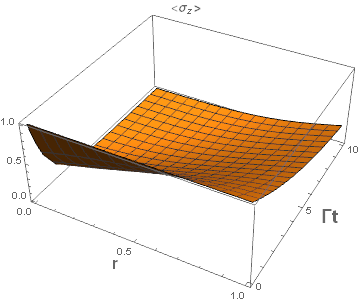}
		\includegraphics[width=6.3cm]{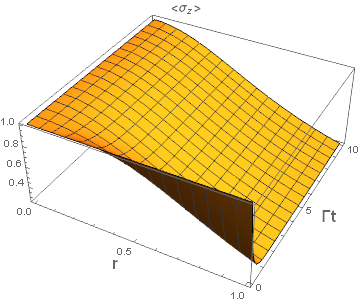}
		\caption{Left (right) figure: {{$ \langle\hat\sigma_{z}\rangle_{ss}$} in terms of the scaled time {$\Gamma t$} and {$r$} without (with) {\small{\mbox{RWA}}} for {$\theta=0$}}.}
	\end{figure}\\
	In Fig. 11 we plotted {$ \langle\hat\sigma_{z}\rangle_{th }$} in terms of the scaled time {$\Gamma t$} and {$n_{th}$} without (with) {\small{\mbox{RWA}}}. We observe in without {\small{\mbox{RWA}}} by increasing in {$\Gamma$}{$t$} and {$n_{th}$} damping is increased. But we observe with {\small{\mbox{RWA}}} by increasing time and {$n_{th}$} damping is decreased.   
	\begin{figure}[!h]
		\includegraphics[width=6.3cm]{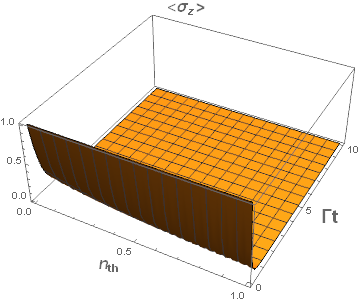}
		\includegraphics[width=6.3cm]{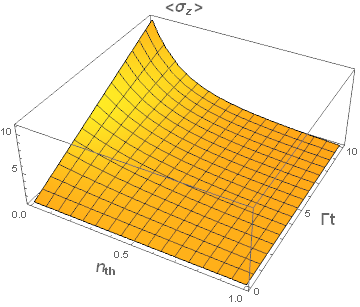}
		\caption{Left (right) figure: {$ \langle\hat\sigma_{z}\rangle_{th}$} in terms of the scaled time {$\Gamma$}{$t$} and {$n_{th}$} without (with) {\small{\mbox{RWA}}}.}
	\end{figure}
	\newpage
	\section{Summary and conclusion}
	In this paper, we investigated the interaction of a two-level atom (and subsequently a single-mode field) with dissipative reservoirs while retaining the CRTs, i.e., without invoking the RWA. We considered two important models of environments, namely thermal and squeezed reservoirs. For both cases, we derived the generalized master (Lindblad) equations governing the time evolution of the reduced density operator of the atom and the field as the systems in our consideration. Since the inclusion of the CRTs significantly modifies the dynamical equations compared with the conventional RWA treatment, one may expect noticeable corrections to the dynamics of physically measurable quantities. To examine the influence of the CRTs, we recalculated several quantities that have previously been studied within the RWA. In particular, we analyzed the time evolution of the expectation values of the Pauli operators, $\langle\hat\sigma_x\rangle$, $\langle\hat\sigma_y\rangle$, and $\langle\hat\sigma_z\rangle$ for the squeezed and thermal reservoirs.
	Our results show that the dynamics of $\langle\hat\sigma\rangle$ are strongly affected by the inclusion of the CRTs. As the squeezing parameter increases, the decay rate of $\langle\hat\sigma_y\rangle_{ss}$ increases in the absence of the RWA, whereas it decreases significantly under the RWA. Moreover, for a thermal reservoir, the exact dynamics without RWA  predicts that $\langle\hat\sigma_y\rangle_{th}$ decays to zero after a finite time for all considered average thermal photon numbers, while under the RWA it instead approaches a nonzero steady state value. For $\langle\hat\sigma_z\rangle_{ss}$, increasing the squeezing parameter leads to a faster decay in both thermal and squeezed reservoirs when the CRTs are retained. In addition, the dependence on the thermal photon number is qualitatively different in the two approaches. Without the RWA, increasing the average thermal photon number slows down the decay of $\langle\hat\sigma_z\rangle_{th}$, whereas under the RWA it accelerates the decay.\\
	These findings clearly demonstrate that the CRTs can produce not only quantitative corrections, but also qualitative changes in the dissipative dynamics of open quantum systems. At least for the observables investigated in this work, neglecting the CRTs may lead to significantly different physical predictions. We therefore expect that the inclusion of the CRTs will play an important role in the accurate description of a broad range of dissipative quantum phenomena, particularly in regimes where the atom–reservoir coupling or the driving strength is sufficiently strong wherein the RWA is no longer valid.
	
	
	\newpage
	\section*{}
	\bibliographystyle{vancouver} 
	\bibliography{ELSRefBank}

\end{document}